\documentclass[prl,twocolumn,showpacs,amsmath]{revtex4}
\usepackage{tabularx}
\usepackage{graphicx}
\usepackage{bm}
\usepackage{amsfonts}
\usepackage{amsmath}
\usepackage{exscale}
\usepackage{amsbsy}

\newcommand{\pder}[2]{\frac{\partial{#1}}{\partial{#2}}}

\newcommand{\av}[1]{\left\langle{#1}\right\rangle}

\def\be{\begin{equation}}
\def\ee{\end{equation}}
\def\ba{\begin{eqnarray}}
\def\ea{\end{eqnarray}}

\def\rr{{\bf r}}

\def\ve{\varepsilon}
\def\D{\Delta}

\def\Tr{\textrm{Tr}}

\def\t{\theta}

\begin{document}

\title{Conductance of d-wave superconductor/normal metal/d-wave superconductor junctions}

\date{\today}

\author{D. A. Pesin, A. V. Andreev, and B. Spivak}
\affiliation{Department of Physics, University of Washington,
Seattle, WA 98195, USA}
\begin{abstract}
We develop a theory of the conductance of superconductor/normal metal/superconductor junctions
in the case where the superconducting  order parameter has $d$-wave symmetry. At low temperature the conductance is proportional to the square root of the inelastic electron relaxation time in the bulk of the superconductor. As a result it turns out to be much larger than the conductance of the normal part of the junction.
\end{abstract}
\pacs{74.45.+c, 74.50.+r, 74.20.Rp}
\maketitle
At small voltages $V$, the  current $I$ through a superconductor/normal
 metal/superconductor (SNS) junction can be written as
\begin{equation}
I=J(\phi) +G(\phi)V.
\label{current}
\end{equation}
Here  $\phi$ is the phase difference between the superconductors, and $J(\phi)$
 is a periodic function of $\phi$ with  period $2\pi$. The time dependence of
  $\phi$ is given by the Josephson relation:
\begin{equation}\label{eq:josephson}
\frac{d \phi}{d t}=2eV
\end{equation}

The first term in Eq.~(\ref{current}), describing the Josephson current, has been  the subject of intensive experimental and theoretical studies over several decades, since the discovery of the Josephson effect. The second term, representing the dissipative current, has attracted relatively little attention, both on the theoretical and experimental sides. In the context of SNS junctions with $s$-wave symmetry of the order parameter in the leads, this problem  was considered theoretically in~\cite{ZaikinJETP,Kopnin,VolkovZaitsev,Aslamazov,Lempitskii,Zhou,Pannetier}, and experimentally in~\cite{Lehnert}. The common result of these works is that at low temperatures the conductance of the system is proportional to the inelastic relaxation time in the normal metal. A theory of conductance of a junction in the case when the superconducting order parameter  has $d$-wave symmetry has not been developed. In this paper we show that at low temperatures $T$ the conductance, $G_{DND}$, is proportional to the square root of the energy relaxation time in the bulk of superconductor, $\tau_{in}$, and that it is much larger than the conductance $G_{N}$ of the normal piece of the junction.

The origin of the leading low-temperature contribution to $G_{DND}$ is similar to the Debye relaxation mechanism
in dielectrics, or the Mandelstam-Leontovich mechanism for sound absorption in liquids with internal degrees
of freedom \cite{LLvol6}. Due to the proximity effect the single particle density of states in the normal
region, $\nu_N(\ve, \phi)$, becomes $\ve$ and $\phi$-dependent. Here $\ve$ is the energy of
quasiparticle. According to Eq.~(\ref{eq:josephson}), $\nu_{N}(\ve,\phi(t))$ changes in time.
In the adiabatic approximation the electron population follows the motion of the levels. As a
 result, the quasiparticle distribution function becomes nonequilibrium. Its relaxation leads to
  the entropy production, and therefore contributes to the conductance:
\begin{equation}\label{eq:entropy_conductance}
T\dot{S}=V^{2}G_{DND}.
\end{equation}

The equation for the entropy production reads
\begin{equation}\label{eq:entropy production}
\dot{S}=4\int d\rr\int^{\infty}_{0} d \varepsilon\,
\frac{\nu\nu_0(\rr)}{(1-f^2_{th})}\left[\frac{D_{ij}}{\nu}\frac{\partial
f}{\partial x_i}\frac{\partial f}{\partial
x_j}+\frac{(f-f_{th})^{2}}{\tau_{in}(\rr)}\right],
\end{equation}
where $f(\ve,\rr,t)$ is the quasiparticle distribution function,  $f_{th}(\ve,\rr,t)= \tanh{\ve/2T}$ is
 the equilibrium distribution function, $\tau_{in}$ is the inelastic relaxation time, $D_{ij}(\ve)$ is the diffusion coefficient, and $\nu$ is the reduced density of states, measured in the units of the normal
state density of states, $\nu_{0}$. We assume that the latter is the same in both metals.

The kinetic equation  describing the dynamics of quasiparticles has the following form:
\begin{widetext}
\begin{eqnarray}\label{eq:finalkineq}
\nu\pder{f}{t}-\frac{\partial}{\partial {x_{i}}}D_{ij}({\bf r})\frac{\partial
f}{\partial {x_{j}}}-\frac{\partial f_{th}}{\partial
\varepsilon}\int_{-\infty}^{\ve}d\ve' \frac{\partial }{\partial t} \nu({\bf
r}, \ve';\phi(t))=
-\frac{\nu (f-f_{th})}{\tau_{in}(\rr)}.
\end{eqnarray}
\end{widetext}
The diffusion form of  Eq.~(\ref{eq:finalkineq}) is valid both inside the normal metal region if $L_N\gg l_{N}$, and in the bulk of $d$-wave superconductors at distances much larger than $l_D$. Here  $L_{N}$ is the length of the normal metal part of the junction, $l_{N,D}$ are the the elastic mean free paths in the normal metal and $d$-wave superconductor, respectively.

The last term in the left hand side of Eq.~({\ref{eq:finalkineq}})
describes the generation of a nonequilibrium quasiparticle distribution function due to the motion of the energy levels. Qualitatively, the origin of this term can be illustrated using the example of a system of discrete Andreev levels labeled by their energy. Due to conservation of the number of levels, their motion is described by a continuity equation
\begin{eqnarray}\label{eq:flow_energy space}
  \pder{N(\ve)}{t}+\pder{j_\ve}{\ve}=0,
\end{eqnarray}
where $j_\ve$ is the ``current of levels'' in energy space, and $N(\ve)$ is the total
 density of states at energy $\ve$. If a level is occupied with a probability $\tilde{f}$,
its motion causes a change in the occupation numbers of quasiparticles.
The rate of change of the total number of quasiparticles at given energy, $N\tilde{f}$, is determined by a continuity equation with a current $\tilde{f}j_\ve$:
\begin{equation}
  \pder{(N\tilde{f})}{t}+\pder{(\tilde{f}j_\ve)}{\ve}=0.
\end{equation}
Then using Eq.~(\ref{eq:flow_energy space}) one can obtain Eq.~(\ref{eq:finalkineq}).

In what follows, we assume that the last term in the left hand side of Eq.~(\ref{eq:finalkineq}), describing the  generation of a nonequilibrium distribution of quasiparticles, is nonzero only in the normal part of the junction. The reason is the following.
At small $T\ll \Delta_{0}$, where $\D_0$ is the amplitude of the order parameter, the voltage drop takes place mainly in the normal metal region. Indeed, the majority of quasiparticles incident on the normal metal-superconductor boundary experience Andreev reflection.
A certain  small fraction of the quasiparticles incident on the boundary in the direction parallel to the node in the quasiparticle spectrum in superconductor can penetrate the boundary.   The part of the distribution function of these quasiparticles  (the distribution function $f_{1}$ in the notation of Ref.~\cite{LO}), which is responsible for the imbalance of electron and hole populations, as well as for
penetration of the electric field into the superconductor, decays away from the boundary on a length scale $L_E $, which is of the order of the elastic mean free path in the superconductor, $L_E \approx l_D\ll L_N$ ~\cite{Artemenko,Choi}.
This is the reason why we take into account  only the electron-hole symmetric part of the distribution function (the distribution function $f$  of Ref.~\cite{LO}) in Eqs.~(\ref{eq:entropy production}) and~(\ref{eq:finalkineq}).

Generally speaking, near the superconductor-normal metal boundary the system does not have $d$-wave symmetry. As a result,  an $s$-component of the anomalous Green function is generated. At distances larger than $l_N$ into the normal metal only this component survives and gives rise to the $\phi$-dependence of the density of states. To describe this effect in the normal metal piece of the junction we use the Usadel equations for angles $\t(\ve,\rr)$ and
 $\chi(\ve,\rr)$, parameterizing the retarded Green function~\cite{Usadel,RammerSmith,Belzig}:
\begin{eqnarray}
&&\frac{D_{N}}{2}\frac{\partial^2\theta}{\partial\rr^2}+i\ve \sin \theta-\frac{D_N}{4}\left(\pder{\chi}{\rr}\right)^2\sin{2\t}=0,\nonumber\\
&&\pder{}{\rr}\left(\pder{\chi}{\rr}\sin^2\t\right)=0,
\label{eq:Usadel}
\end{eqnarray}
where $D_N$ is the diffusion coefficient in the normal metal, and we dropped the time derivative, which is justified provided that $eV\ll E_{c}=D_{N}/L_{N}^{2}$.
The normal and anomalous retarded Green functions, and the local density of states in the metal can be expressed as $g^{R}(\ve, {\bf r})= \cos \theta(\ve, {\bf r})$, $f^{R}(\ve, {\bf r})= -i\sin \theta(\ve,{\bf r})e^{i\chi}$, and $\nu({\bf r},\ve)=\textrm{Re}\,\cos \theta(\ve, {\bf r})$. Elastic electron scattering suppresses $d$-wave superconductivity. Thus we consider the case $\xi_{D}\ll l_D$, where $\xi_{D}$ is the superconductor zero temperature coherence length.

\begin{figure}
\begin{center}
\includegraphics[scale=1.3,bb=238 652 370 722]{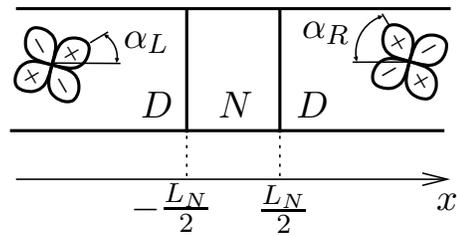}
\caption{Schematic picture of a $d$-wave superconductor (D)/normal metal (N)/$d$-wave superconductor junction.}
\label{fig:geometry}
\end{center}
\end{figure}

Boundary conditions for Eqs.~(\ref{eq:Usadel}) in the case of a contact between a $d$-wave superconductor and a diffusive normal wire were derived in Ref.~\cite{Nazarov}. They depend on the angle $\alpha$ between the crystal axis of the superconductor and the normal to the $ND$-boundary (see Fig.~\ref{fig:geometry}). In the case  $\alpha_L=\alpha_R=0$, for a high transparency boundary, and at $T, E_{c}\ll \Delta_{0}$, the boundary conditions reduce to
\begin{eqnarray}\label{eq:Usadel_boundary conditions}
&&\theta(\ve,x=\pm L_N/2)=\t_B\equiv \arcsin(\sqrt{2}-1)\approx 0.43,\nonumber\\
&&\chi(\ve,x=\pm L_N/2)=\pm \frac{\phi}{2}.
\end{eqnarray}

In the case of $s$-wave superconductivity the system of Eqs.~(\ref{eq:finalkineq}) and~(\ref{eq:Usadel}) was derived in \cite{LO}. In the $d$-wave case, and for energies larger than the zero-energy scattering rate $\Gamma$ (defined below), Eq.~(\ref{eq:finalkineq}) follows from the Boltzmann  kinetic equation for the quasiparticle distribution function. To see that Eq.~(\ref{eq:finalkineq}) is valid independently of the relation between $\ve$ and $ \Gamma$ we illustrate its derivation in the Born limit of impurity scattering. To this end we use the quasiclassical equation for the Keldysh Green function $\check{g}(\ve,t)$, which is a $4\times4$ matrix in the Keldysh and Gorkov-Nambu spaces~\cite{LO}, neglecting  inelastic scattering:
\begin{eqnarray}\label{eq:qcKeldysh}
\frac{1}{2} [\tau_3,\dot{\check{g}}]_{+}+\vec{v}_F\nabla \check{g}-i[\ve\tau_3+i\tau_2\D_{\bf k},\check{g}]+\frac{1}{2\tau_D}
 [\langle{\check{g}}\rangle,\check{g}]_{-}=0,
\end{eqnarray}
where $[\ldots,\ldots]_{\mp}$ denotes commutator and anticommutator, $\tau_{2,3}$ are the Pauli matrices in the Gorkov-Nambu space, $\D_{{\bf k}}$ is a uniform $d$-wave order parameter corresponding to the direction ${\bf k}$ on the Fermi surface, $\tau_D$ and $v_{F}$ are the normal state elastic mean free time and Fermi velocity in the $d$-wave superconductor. Introducing the generalized distribution functions $f$ anf $f_1$ via $\hat{g}^K=(\hat{g}^R-\hat{g}^A)f+(\hat{g}^R\tau_3-\tau_3\hat{g}^A)f_1$ ~\cite{LO},
Eq.~(\ref{eq:qcKeldysh}) gives two equations for the latter:
\begin{widetext}
\begin{eqnarray}\label{eq:kineq}
&&u\pder{f}{t}+u \vec{v}_F\nabla f_1=-\frac{u}{\tau_D}(\av{u}f-\av{uf})+\frac{v}{\tau_D}(\av{v}f-\av{vf})\nonumber\\
&&u\vec{v}_F\nabla f-2i\D_{\bf k} wf_1=-\frac{u}{\tau_D}(\av{u}f_1-\av{uf_1})+\frac{w}{\tau_D}(\av{w}f_1-\av{wf_1}),
\end{eqnarray}
\end{widetext}
where $\langle\ldots\rangle$ stands for the average over directions on the Fermi surface, and the quantities
\ $u,v$ and $w$ are defined by $u=\Tr\,(\hat{g}^R-\hat{g}^A)/4$, $v=-i\Tr\,\tau_2(\hat{g}^R-\hat{g}^A)/4$, and $w=-i\Tr\,\tau_2(\hat{g}^R+\hat{g}^A)/4$. The time derivative of $f_1$ in the second of Eqs.~(\ref{eq:kineq}) can be neglected if we do not consider generation of branch imbalance in the superconductor. Expressing $f_1$ via $\vec{v}_F\nabla f$ using the second of Eqs.~(\ref{eq:kineq}), and averaging the first one  over direction of momentum, we obtain Eq.~(\ref{eq:finalkineq}), with the following expressions for the diffusion coefficient and the density of states, valid in both Born and unitary limits:
\begin{eqnarray}\label{eq:DiffCoeff}
    D_{ij}(\ve)&=&\frac{v^2_{F}}{4\textrm{Re}t\textrm{Im}t}\left\langle \hat{k}_i
    \hat{k}_j\textrm{Re}\frac{|t|^2+t^2-2\D^2_{\bf k}}{\sqrt{t^2-\D^2_{{\bf k}}}}\right\rangle,
\end{eqnarray}
\begin{equation}
\nu_D(\ve)=\left\langle\textrm{Re}\frac{t}{\sqrt{t^2-\D^2_{\bf k}}}\right\rangle.
\label{DOS}
\end{equation}
The quantity $t(\ve)$ determines the retarded Green function, $\hat{g}^R=(\tau_3t+i\tau_2\Delta_{\bf k})/\sqrt{t^2-\Delta^2_{\bf k}}$, and
is defined by the following equation:
\begin{eqnarray}\label{eq:t}
     t(\ve)&=&\ve+\frac{i}{2\tau_D}\left\langle\frac{t(\ve)}{\sqrt{t^{2}(\ve)-\D^2_{\bf k}}}\right\rangle^\beta,
\end{eqnarray}
where $\beta=\pm 1$ corresponds to the Born and unitary limits of impurity scattering, respectively. The structure of Eqs.~(\ref{eq:DiffCoeff}),~(\ref{DOS}), and~(\ref{eq:t}) is very similar to that appearing in the theory of the electronic thermal conductivity of $d$-wave superconductors~\cite{Lee, Mineev}.

Let us consider the geometry of the junction shown in Fig.~\ref{fig:geometry}. We consider the $d$-wave superconductor to be a stack of two-dimensional layers~\cite{GrafLayers}. The planes of the layers are parallel to the axis of the junction. We also assume that within each layer the order parameter has $d_{x^2-y^2}$ symmetry, with order parameter given by $\D_{\bf k}=\D_0\cos[2(\varphi-\alpha_{L,R})]$ in the left and right leads, respectively. In this case the diffusion coefficient, Eq.~(\ref{eq:DiffCoeff}), for diffusion along the layers reduces to a scalar, $D_{ij}(\ve)=D_D(\ve)\delta_{ij}$. Using Eqs.~(\ref{eq:DiffCoeff}) and~(\ref{eq:t}) we get
\begin{equation}
D_{D}(\ve)=\left\{\begin{array}{l}
  \frac{v^2_F\tau_D}{4}\eta^2(\ve),\,\,\, |\ve| \gg\Gamma,  \\
  \frac{v^2_F}{2\pi \D_0},\,\,\,\, |\ve| \ll\Gamma,
\end{array}\right.
\label{diffCoefD}
\end{equation}
where $\eta(\ve)=1$ in the Born limit, and $\eta(\ve)\approx[2|\ve|\ln(\D_0/|\ve|)]/\pi\D_0$ in the unitary limit. For the density of states we obtain~\cite{GorkovKalugin}
\begin{equation}
\nu_{D}(\ve)=\left\{\begin{array}{l}
  \frac{|\ve|}{\D_0},\,\,\, |\ve| \gg\Gamma,  \\
  \frac{2\Gamma}{\pi\D_0}\ln\frac{\D_0}{\Gamma},\,\,\,\, |\ve| \ll\Gamma.
\end{array}\right.
\label{DoSD}
\end{equation}
In the above equations the zero-energy scattering rate, $\Gamma$, is given by \begin{equation}
\Gamma=\textrm{Im}t(\ve\to 0)\approx\left\{\begin{array}{l}
  \D_0 e^{-\pi\D_0\tau_D},\,\,\,\,\textrm{Born limit},\\
  \sqrt{\frac{\pi\D_0}{2\tau_D\ln(\D_0\tau_D)}},\,\,\,\,\textrm{unitary limit}.
\end{array}\right.
\end{equation}

We now turn to Eq.~(\ref{eq:kineq}). Formally, it should be supplemented with boundary conditions at the $ND$ boundaries. However, provided that the temperature is low enough, the time a quasiparticle spends in the normal region is small compared to $\tau_{in}$.  Therefore, the third term in the left hand side of Eq.~(\ref{eq:finalkineq}) can be substituted by $I_N(\ve,t)\delta(x)$, where
\begin{equation}\label{eq:source}
  I_N(\ve,t)=L_N\pder{f_{th}}{\ve}
\int_{-\infty}^{\ve}d\ve' \frac{\partial \bar{\nu}_N(\ve',t)}{\partial t},
\end{equation}
and $\bar{\nu}_N(\ve,t)$ is the density of states in the normal part of the junction averaged over its length,
\begin{equation}
\bar{\nu}_N(\ve,t)=\frac{1}{L_N}\int^{L_N/2}_{-L_N/2}dx\, \nu_N(x,\ve,t).
\end{equation}

Solution of Eqs.~(\ref{eq:Usadel}),~(\ref{eq:Usadel_boundary conditions}), analogous to that presented in Refs.~\cite{Pannetier, Blatter}, yields the following $\ve$ and $\phi$ dependence of the density of states in the normal metal region
\begin{eqnarray}\label{eq:nuNormal}
\bar{\nu}_N&=&\left\{\begin{array}{l}
 \cos\t_B, \,\,\,\,\phi=0,\,|\ve|\ll E_{c},\\
 \sin\t_B/\t_B,\,\,\,\,\phi=\pi
,\,|\ve|\ll E_{c},\\
\approx 1,\,\,\,\,|\ve|\gg E_c.
\end{array}\right.
\end{eqnarray}

The solution of Eq.~(\ref{eq:finalkineq}) is
\begin{equation}\label{eq:diffeq solution}
  f-f_{th}=I_N\frac{L_\ve}{2D_D}e^{-|x|/L_\ve},\,\,\,\,
L_\ve=\sqrt{\frac{D_D\tau_{in}}{\nu_D}}.
\end{equation}

Substituting Eq.~(\ref{eq:diffeq solution}) into Eq. (\ref{eq:entropy production}),
 at $eV\ll 1/\tau_{in}$ we obtain the conductance
\begin{eqnarray}\label{conductance}
G_{DND}(\phi)&\approx&2 G_N \frac{\sqrt{E_c\tau_{in}}}{E^{2}_c}\nonumber\\ &\times&\int^{\infty}_{0} d \ve\,
\pder{f_{th}}{\ve}\sqrt{\frac{D_N}{\nu_D D_{D}}}
\left(\int_{-\infty}^{\ve}d\ve' \frac{\partial \bar{\nu}_N(\ve',\phi)}{\partial \phi}\right)^2.
\end{eqnarray}
Eq.~(\ref{conductance}) is the main result of this paper, which shows that
the low-temperature conductance is proportional to the square root of the inelastic relaxation time in the bulk of the $d$-wave superconductor.

At low temperature the energy relaxation in $d$-wave superconductors is determined by  electron-phonon and electron-electron interactions.
We would like to mention, however, that regardless of the relation between the electron-electron, $\tau_{e-e}$, and electron-phonon, $\tau_{e-ph}$, relaxation times, the conductance remains proportional to $\sqrt{\tau_{e-ph}}$. The reason is that electron-electron scattering processes conserve energy. Thus if $\tau_{e-e}\ll \tau_{e-ph}$, at short time scales electron-electron scattering processes establish an equilibrium form of the distribution function with a non-equilibrium value of temperature, which then relaxes via electron-phonon processes.

In the framework of BCS theory the electron-phonon relaxation rate in a $d$-wave superconductor at  values of $T$ that are not too small has the same order of magnitude as in the normal metal $\tau_{e-ph}\propto \Theta_{D}^{2}/T^{3}$, where $\Theta_{D}$ is the Debye energy. The reason is that the emission of phonons is not associated with a significant change in direction of the electron momentum. We  note that in the context of high-$T_c$ materials the issue of the value and the $T$ dependence of $\tau_{in}$ has not yet been settled~\cite{ScalapinoTAU}.

According to Eqs.~(\ref{eq:josephson}) and~(\ref{conductance}) the conductance of the junction $G_{DND}(\phi(t))$ is a periodic function of $\phi$ with  period $2\pi$, or a periodic function of time with period $\pi/eV$.
The amplitude of the oscillations is of the order of the average over the period of the oscillations conductance, $\overline{G_{DND}}$. The value of $\overline{G_{DND}}$ depends on the ratio between $E_{c}$, $T$, and $\Gamma$:
\begin{equation}\label{eq:conductanceGamma}
  \overline{G_{DND}}\sim G_{N}
\t^4_B\sqrt{\D_0\tau_{in}}
\left\{
\begin{array}{l}
\frac{1}{\eta(E_c)}\frac{E_c}{T};\,\,\,\,T\gg E_c\gg \Gamma,\\
\frac{1}{\eta(\Gamma)}\sqrt{\frac{E^3_c}{T^2\Gamma}};\,\,\,\,T,\Gamma\gg E_c,\\
\frac{1}{\eta(T)}\left(\frac{T}{E_c}\right)^{3/2};\,\,\,\,E_c\gg T\gg \Gamma,\\
\frac{1}{\eta(\Gamma)}\frac{T^2}{\sqrt{E_c^3\Gamma}};\,\,\,\,E_c,\Gamma\gg T.
\end{array}
\right.
\end{equation}
Here the function $\eta(\ve)$ is defined after Eq.~(\ref{diffCoefD}). We also assumed that the normal state diffusion coefficients in the metals are the same. As a function of $\alpha_{L,R}$, $G_{DND}$, Eq.~(\ref{conductance}), has a maximum given by Eq.~(\ref{eq:conductanceGamma}) at  $\alpha_{L}=\alpha_{R}=0$,
and vanishes at $\alpha_{L}=\alpha_{R}=\pi/4$. In the latter case the conductance of the junction can be calculated in a way similar to calculations of the conductance of an $ND$ junction, as in Ref.~\cite{Nazarov}, and in that case $G_{DND}<G_{N}$.

We note that Eqs.~(\ref{conductance}) and~(\ref{eq:conductanceGamma}) are valid at low voltages, $eV<1/\tau_{in}$. At $eV>1/\tau_{in}$, the same considerations lead to a non-analytic $V$-dependence of the $\overline{G_{DND}}(V)\propto V^{-1/2}$, which can be obtained from Eq.~(\ref{eq:conductanceGamma}) by the substitution $\tau_{in}\to 1/2eV$.

We would like to contrast our result, Eq.~(\ref{eq:conductanceGamma}), with that for an SNS junction with $s$-wave pairing symmetry in the leads \cite{ZaikinJETP,Kopnin,Lehnert,Aslamazov,VolkovZaitsev,Lempitskii,Zhou,Pannetier}, in which the low-temperature
conductance is proportional to the inelastic relaxation time in the normal metal,  $G_{SNS}\propto \tau_{in}$. The origin of the difference with the $s$-wave case is that due to Andreev reflection the nonequilibrium quasiparticles cannot escape the normal region.

The possibility of observing the effect discussed above is limited by the
condition $eV<1/\tau_{in}$. At low temperatures $\tau_{in}$ is long, and this condition is quite restrictive for junctions with leads made out of low-$T_c$ superconductors. For junctions made out of high-$T_c$ superconductors the value of $\tau_{in}$ can be much smaller, and the above requirement on voltages is much weaker. The values of the relaxation times in high-$T_c$ superconductors are not well established. If we estimate $\tau_{in}\sim 10$ps at  $T\sim 20K$ in YBaCuO~\cite{Tau}, we get the region of applicability of our theory  $V<100 \mu V$. We also note that the length of the junction is limited by the diffusion length $\sqrt{D_N \tau^N_{e-e}}$ that is determined by the electron-electron relaxation time in the normal metal, $\tau^N_{e-e}$.

The obtained results are valid well below the critical temperature, $T\ll T_c$. Near $T_c$, Andreev reflection at an $ND$ boundary, which gives rise to the results presented above, becomes inefficient. Also the penetration length of the electric field into superconductor diverges as $T\to T_{c}$, and the resistance of the junction can be calculated in the same way as a resistance of an $NS$ boundary~\cite{Artemenko}.

\acknowledgments

The authors would like to thank C.W.J. Beenakker, Yu.V. Nazarov and A.F. Volkov for useful discussions. This work was supported by the DOE Grant No. DE-FG02-07ER46452, and the NSF Grant No. DMR-0704151.

\end{document}